\documentclass[twocolumn,prb,aps,epsf]{revtex4}

\usepackage{graphicx}
\usepackage{dcolumn}
\usepackage{bm}
\usepackage{hyperref}

\begin{document}

\title{Positive and negative Coulomb drag in vertically-integrated one-dimensional quantum wires}

\author{D. Laroche$^{1,2}$, G. Gervais$^{1}$, M.P. Lilly$^{2,*}$ and J.L. Reno$^{2}$ }

\affiliation{$^{1}$Department of Physics, McGill University, Montreal, H3A 2T8, CANADA}

\affiliation{$^{2}$Center for Integrated Nanotechnologies, Sandia National Laboratories, Albuquerque, NM 87185 USA}

\affiliation{$^{*}$ e-mail : mplilly@sandia.gov}

\vspace*{-7mm}
\author{Nature Nanotechnology \textbf{6}, 793 - 797 (2011). Published online: 30 October 2011. \\ {http://www.nature.com/nnano/journal/v6/n12/full/nnano.2011.182.html}}

\maketitle


\textbf{Electron interactions in and between wires become increasingly complex and important as circuits are scaled to nanometre sizes, or employ reduced-dimensional conductors \cite{Screening} like carbon nanotubes \cite{cnt1,cnt4, cnt-proxy1, cnt-proxy2, cnt-proxy3}, nanowires \cite{nanowires1, nanowires3, surface_effects2, nanowires-proxy2} and gated high mobility 2D electron systems \cite{Debray, Yamamoto, Ed1}.  This is because the screening of the long-range Coulomb potential of individual carriers is weakened in these systems, which can lead to phenomenon such as Coulomb drag: a current in one wire induces a voltage in a second wire through Coulomb interactions alone. Previous experiments have observed electron drag in wires separated by a soft electrostatic barrier $\gtrsim$ 80 nm \cite{Yamamoto}.  Here, we measure both positive and negative drag between adjacent vertical quantum wires that are separated by $\sim$ 15 nm and have independent contacts, which allows their electron densities to be tuned independently.  We map out the drag signal versus the number of electron subbands occupied in each wire, and interpret the results in terms of momentum-transfer and charge-fluctuation induced transport models.  For wires of significantly different subband occupancies, the positive drag effect can be as large as $25\%$. }

Our report addresses the fundamental issues as to what one might expect when coupling quantum circuits in close proximity at the nanoscale. As the transport channel size is reduced towards the one-dimensional limit, charge flow across the channel becomes increasingly dominated by quantum processes.  Due to the long-range nature of the Coulomb potential, coupling two quantum circuits in close proximity (separated by a hard barrier of width $d$) may have profound effects on the current flow and on the equilibrium charge distribution in one wire when current is driven in another wire. First, when $d$ is only a few nanometres, tunneling may occur between the two circuits and induce a current. This tunneling current is strongly suppressed with increasing $d$. Even after tunneling becomes negligible, a non-zero potential across one circuit may appear when current flows in the second circuit as a result of Coulomb interactions.  This resulting `drag signal' depends critically on the inter-circuit separation, the electronic wire density, and electron-electron interactions in the wire. We have fabricated a device that allows for a large degree of electrical control and tunability between two one-dimensional (1D) quantum circuits, thereby providing us with the platform to study in detail the Coulomb drag signal emanating between two quantum wires separated by only a few tens of nanometre.

The ability to independently control the density of each component in a quantum circuit is an important asset needed to thoroughly characterize the interactions between closely spaced low-dimensional structures.  Designs allowing such independent characterization have successfully been implemented in electron-electron \cite{Gramilda}, hole-hole \cite{Jorger} and electron-hole \cite{Seamons} bilayer systems and have led to a great understanding of the interaction mechanisms in two dimensions.  Several experiments have been realized in strongly coupled one-dimensional systems consisting of carbon nanotubes \cite{cnt-proxy1, cnt-proxy2, cnt-proxy3}, nanowires \cite{nanowires3, nanowires-proxy2} or quantum wires networks coupled vertically \cite{Auslaender2, Ed1, Pepper} or laterally \cite{Debray, Yamamoto}.  However, with the exception of laterally coupled quantum wires, none of these one-dimensional networks have been realized with independent electrical contacts and independently tunable density allowing for the measurement of each system component. This is in contrast to our vertically-coupled electrical design where each wire has its own ohmic contacts and capacitively coupled gates.

There are two distinct approaches that can be taken when designing quantum wires coupled by proximity.  The first approach is to couple the wires {\it laterally} using an electrostatic gate to separate both circuits.  Such design allows fabrication of a quantum structures with independent contacts and tunable density, as first demonstrated by Debray {\it et al.} \cite{Debray}, and subsequently by Yamamoto {\it et al.}\cite{Yamamoto}  The depth of 2D systems in GaAs heterostructures (typically 80 nm or greater) and the fringing fields of surface defined gates impose that the effective barrier between lateral one-dimensional circuits are soft and are no less than 80 nm.  Thus, in order to construct coupled circuits in the 10 nm range, one must couple the circuits {\it vertically}. In this design, a hard barrier is introduced during the material growth process, allowing coupling electrical circuits over distances of only a few nanometres. The price to pay in this approach is the complex fabrication process \cite{EBASE} required for defining quantum wires with independent electrical contacts, as sketched in Fig. 1\textbf{a-d}.

\begin{figure}
\begin{center}
\includegraphics[width = 9cm]{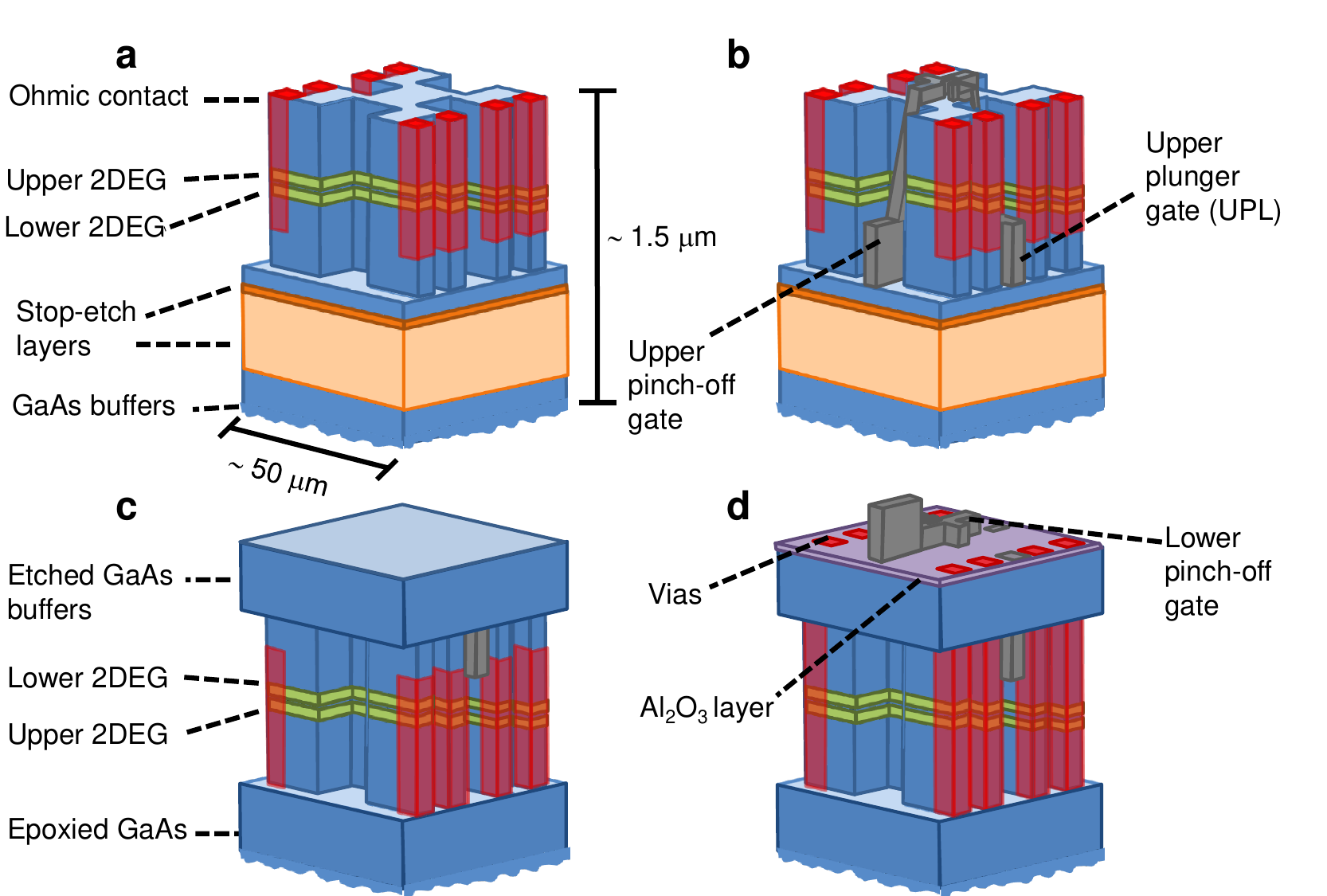} \\
  \caption{\label{figure 1} \textbf{Schematics of the fabrication process of the vertically-coupled quantum circuits.} \textbf{a)} Diagram of the double quantum wires device subsequent to mesa etching, and to Ge-Au-Ni-Au ohmic contacts deposition and annealing.  For visibility purposes, the scale bar in the x-y direction (50 $\mu$m) is dramatically larger than the one in the z-direction (1.5 $\mu$m). \textbf{b)} Diagram showing the deposited upper pinch-off and plunger Ti-Au gates.  The off-mesa section of the gates is patterned using photo-lithography while electron-beam lithography is used to define the on-mesa gates.  \textbf{c)} Diagram after the epoxy-bond-and-stop-etch (EBASE) procedure. Note that due to a flipping process following the new substrate bonding, the upper 2DEG is now at the bottom. Similarly, the upper gates are now buried between the mesa and the epoxied GaAs. The original substrate has been lapped and etched down to $\sim 300$ nm.  \textbf{d)} Diagram showing the final layout of the double quantum wires device after an $Al_{2}O_{3}$ insulating layer is deposited, vias are etched through the device to connect the upper gates and the ohmic contacts to the surface, and another set of Ti-Au split gates is deposited.  More details are presented in the methods section.}\vspace*{-7mm}
\end{center}
\end{figure}

\begin{figure}
\begin{center}
\includegraphics[width = 8cm]{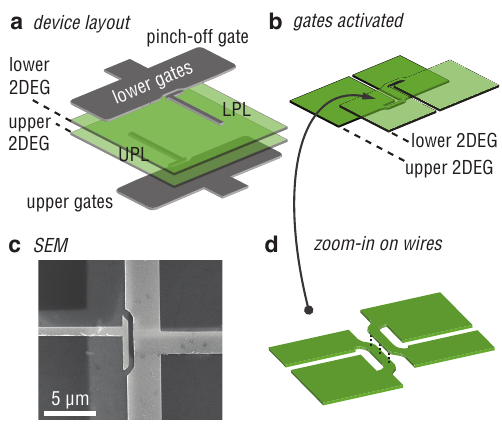} \\
\caption{\label{sketch} \textbf{Split gates design generating the double quantum wire structure.} \textbf{a)} Schematic of the active part of the double quantum wires device.  The EBASE process causes the lower gates and the lower 2DEG to be above the upper gates and 2DEG. \textbf{b)} Schematic of the active part of the device when a suitable bias is applied on all four split gates, effectively coupling both circuits solely through one dimensional regions.  The T-shaped pinch-off gates are simultaneously adjusted to deplete their respective 2DEG, effectively preventing any current flow in the section of the layer underneath (above) the lower wire (upper wire) and creating two independently contacted 2DEGs.  Using the plunger gates, two quantum wires are then formed. \textbf{c)} Scanning electron microscope picture of the device.  The lower plunger (LPL) and pinch-off gates are visible on the surface of the device. \textbf{d)} Zoom-in on the interacting region of the device.  More details are presented in the methods section. }\vspace*{-7mm}
\end{center}
\end{figure}

Fig. 2\textbf{a} shows a schematic of the interacting region of such a device.  This design was modified from previous work \cite{Ed1} so as to enable the measurement of both wires individually, simultaneously and independently.  Applying a suitable voltage on the gates selectively depletes the two-dimensional electron gases (2DEGs) such that two independently contacted quantum wires are created, as shown in Fig. 2\textbf{b} and Fig. 2\textbf{d}.  In this regime, only 1D regions are vertically coupled and parasitic 2D coupling is minimized.  A scanning electron microscope picture of the device is shown in Fig. 2\textbf{c} where $\sim 4.0$ $\mu m$ long and $\sim 0.5$ $\mu m$ wide wires are observed.  It is important for the alignment between the upper and the lower gates to be lesser than, or equal to 30 nm in the direction perpendicular to the wires to ensure a sub 50 nm effective center-to-center distance between the wires.  For the device presented here, this alignment error was less than 25 nm, leading to an effective center-to-center distance between the wires bounded between 33 nm and 41 nm.  Finally, the accuracy of the alignment in the direction parallel to the wires is not as crucial as it only affects the 1D-1D interacting length, which is determined to be 2.8 $\mu m$ long in this device.

\begin{figure}
\begin{center}
\includegraphics[width = 10cm]{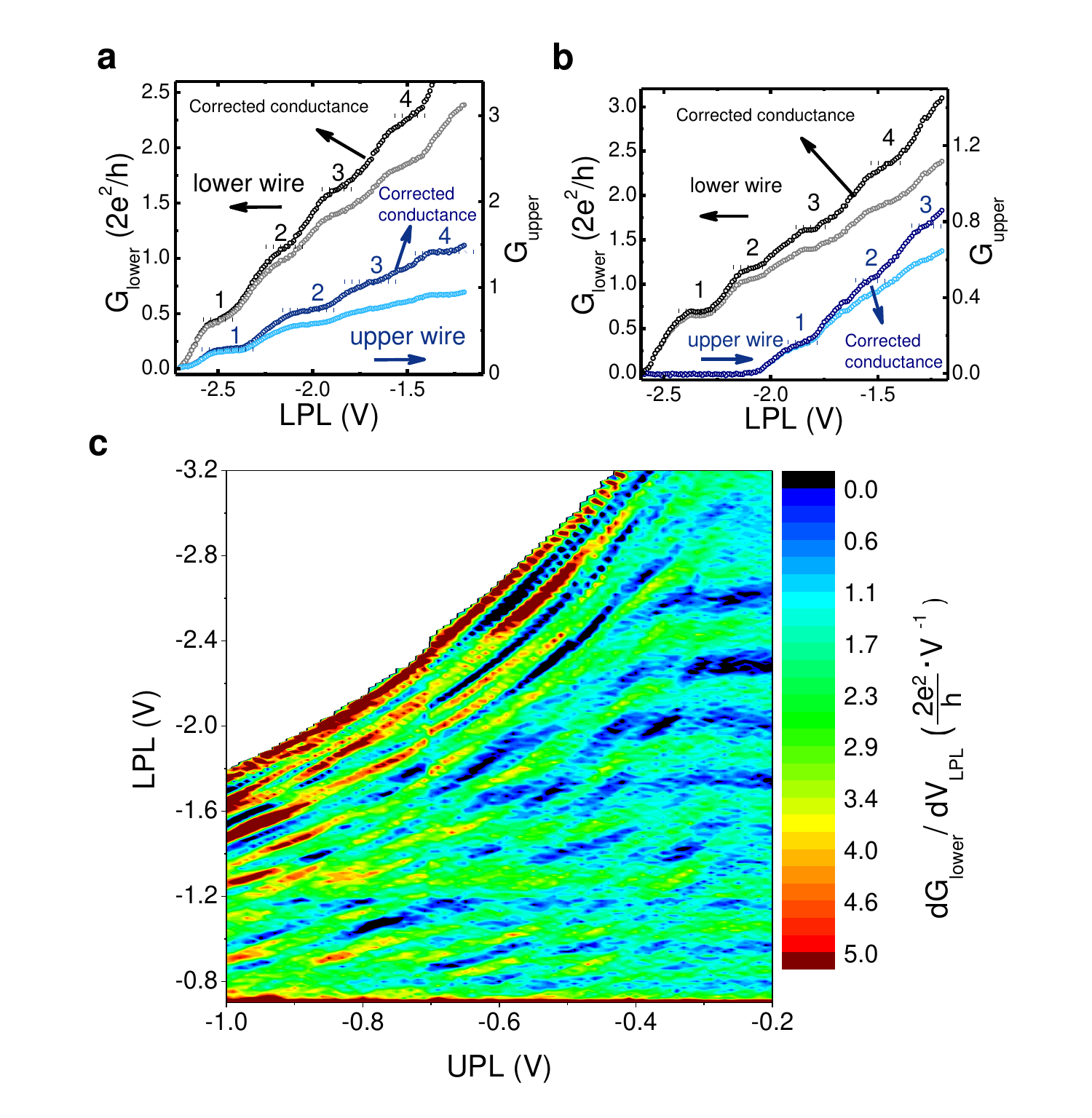} \\
\caption{\label{sketch} \textbf{Characterization of the non-ballistic quantum wires.} Conductance (gray) and corrected conductance (black) in the lower wire (left-axis) and in the upper wire (blue and dark blue curve respectively, right-axis) as a function of LPL voltage for \textbf{a)} fixed UPL = -0.23 V and similar subband occupancies in both wires, and \textbf{b)} fixed UPL = -0.34 V and significantly different subband occupancies in both wires.  For the corrected conductance, 1.25 k$\Omega$ (5.00 k$\Omega$) series resistance was subtracted from the lower (upper) wire conductance. \textbf{c)} Derivative of the lower wire conductance as a function of LPL voltage.  Conductance plateau-like features appear as black and blue stripes in the figure.} \vspace*{-7mm}
\end{center}
\end{figure}

Fig. 3 shows the evolution of the conductance at a temperature of $0.33$ K in each quantum wire as a function of gate voltage.  Fig. 3\textbf{a} was taken at a fixed UPL = -0.23 V where both wires have almost identical 1D subband occupancies for a given LPL voltage. For a fixed UPL = -0.34 V, the wires subband occupancies differ greatly, as depicted in Fig. 3\textbf{b}. These measurements demonstrate that our design allows control of the number of occupied subbands in each circuit.

In the ballistic regime, electron transmission is unhindered and $\sum_{i=1}^{N} T_{i} = N $ in the quantum transport conductance $G = \frac{2e^{2}}{h} \sum_{i=1}^{N} T_{i}$,  while an increase in scattering along the wire causes $\sum_{i=1}^{N} T_{i} < N $ in the non-ballistic regime.  The wires presented here are in the non-ballistic regime and the spacing between the conductance plateau features is less than $\frac{2e^{2}}{h}$.  Correcting for the contact resistance by subtracting a series resistance to the quantum wires yields an even conductance spacing between the plateau-like features, albeit one smaller than $\frac{2e^{2}}{h}$, as shown in Fig. 3{\bf a} and 3{\bf b}. A 1.25 k$\Omega$ (5.00 k$\Omega$) contact resistance was subtracted from the lower (upper) wire conductance.  This higher value of the contact resistance is due to partial depletion of the upper 2DEG when the gates are biased for the drag measurement (see Methods).  An even conductance spacing of plateau-like features at values lower than $\frac{2e^{2}}{h}$ in quasi one-dimensional structures was previously observed \cite{Yacobywire2} and has been found not to affect the 1D nature of the quantum wires. We also show in Fig 3{\bf c} the derivative of the conductance of the lower wire as a function of LPL voltage. Plateau-like features are observed when the derivative approaches zero, appearing as black and blue stripes in this mapping.  This tracking of the plateau-like features, combined with their even conductance spacing, strongly supports the existence of well-defined one-dimensional subbands.

\begin{figure}
\begin{center}
\includegraphics[width = 9cm]{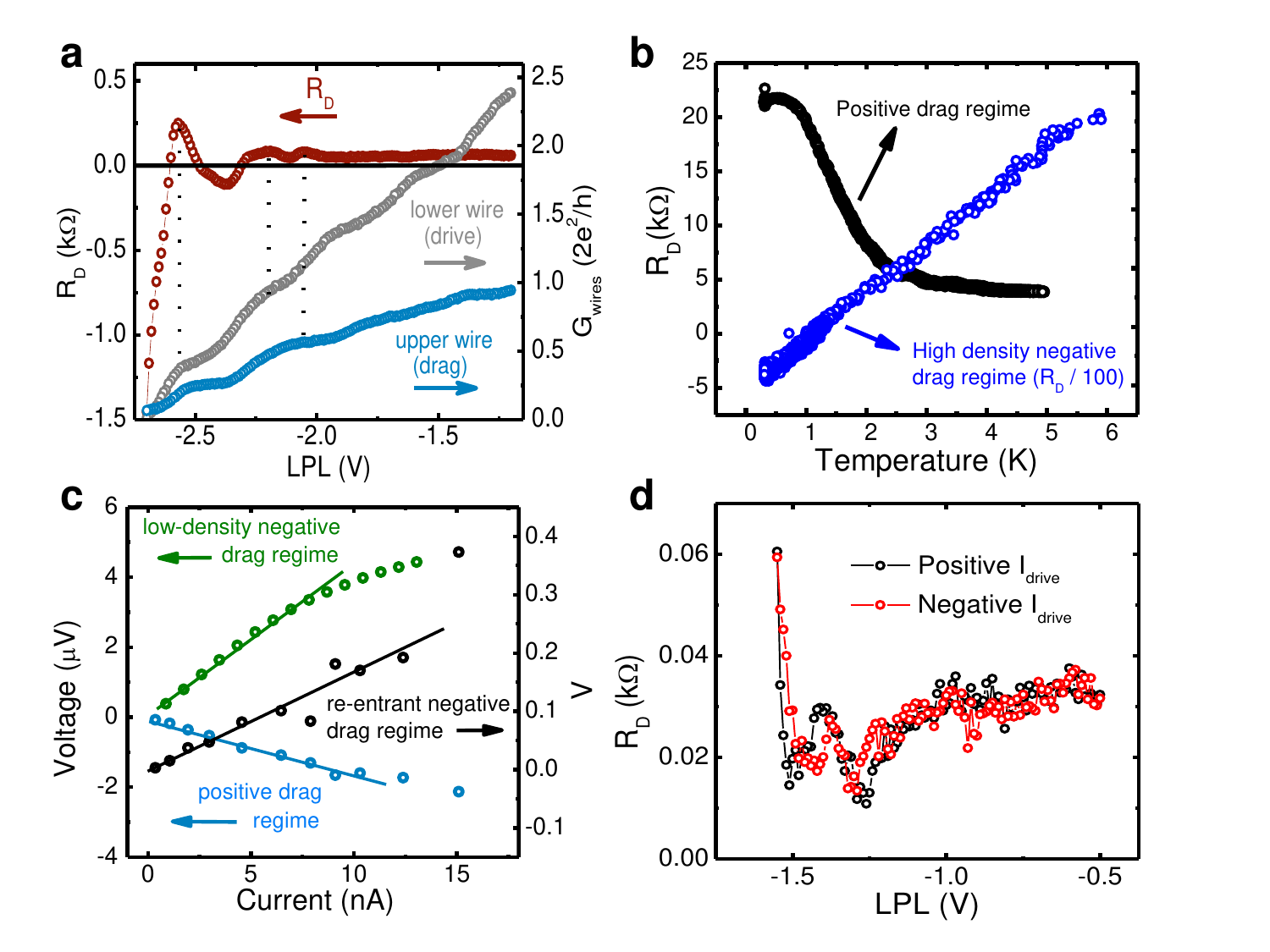} \\
\caption{\label{sketch} \textbf{Drag resistance of the coupled quantum circuits.} \textbf{a)} The drag resistance (red curve, left-axis) is shown as a function of LPL voltage along with the conductance in the drive wire (gray curve, right-axis) and in the drag wire (blue curve, right-axis) for fixed UPL = -0.23 V.  The presence of peaks in the drag resistance concomitant with the opening of 1D channels in either wire is highlighted by doted lines. \textbf{b)} Temperature dependence of the Coulomb drag signal at the peak of the positive drag regime (black curve) and in the high-density negative drag regime (blue curve).  \textbf{c)} Drag voltage as a function of the drive current for the low-density negative drag regime, the positive drag regime and the high-density re-entrant negative drag regime.  In all three regimes, the drag voltage is linear with drive current for $eV_{drive}/K_{b} \lesssim 3 K$.  The current used for the drag measurement (4.5 nA) was always within the linear drag regime.  \textbf{d)} Drag voltage as a function of LPL for both positive (black curve) and negative (red curve) drive currents showing that the signal is independent of the drive current direction. }

\vspace*{-7mm}
\end{center}
\end{figure}


Coupling two independent electrical circuits by proximity may lead to signals in one circuit whose origin emanates entirely from the neighboring circuit such as Coulomb drag. To measure this drag effect, a current $I_{drive}$ is set in one of the (drive) circuits. Under the condition of no current flow, a voltage $V_{drag}$ develops across the second (drag) circuit, defining a drag resistance $R_{D} \equiv - V_{drag} / I_{drive}$ that is a direct probe of electron-electron interactions. Coulomb drag is distinct from rectification and ratchets mechanisms where a voltage develops due to a neighboring current flow whose I-V characteristics are highly non-linear (with respect to $I_{drive}$) and non-symmetric with respect to probe inversion. In contrast, Coulomb drag is an equilibrium phenomena that is linear, invertible with respect to probe symmetry, mutual, and present in ballistic and non-ballistic circuits.

The drag resistance measured in our quantum circuit is shown in Fig. 4\textbf{a}, along with the conductance of each wire. Coulomb drag peaks are observed concomitant with the opening of 1D subbands in either wire (see dotted lines in Fig. 4\textbf{a}).  Momentum matching between both wires can explain the presence of the positive drag peaks when the wires have similar subband occupancies \cite{Debray}, but an enhancement of the electron-hole asymmetry as 1D channels open in the quantum wires \cite{Kamenev}, appears more probable to explain the presence of positive peaks when the wires have different subband occupancies.  In addition, negative Coulomb drag is observed in two clearly distinct regimes : one at low electronic density when the drag wire is close to or beyond depletion, and one at higher electronic density when
$ N_{drag} > 1$. Negative Coulomb drag has been previously observed at low density (\emph{i.e.} for $N < 1$ in both wires)\cite{Yamamoto} and attributed to one-dimensional Wigner crystallization.  While Wigner crystallization could explain the low-density negative drag reported in this Letter, it cannot explain the high-density negative drag.  Negative Coulomb drag has been predicted to occur following a charge-fluctuation induced Coulomb drag model in asymmetric mesoscopic circuits \cite{Kamenev, Sanchez}, but more work is required to assess its consistency over the whole space-phase of 1D Coulomb drag.

We show in Fig. 4\textbf{b} the temperature dependence between $\sim 0.4 K$ and $\sim 6 K$ in both the high-density negative drag and the positive drag regimes. In either case, the drag resistance shows no saturation down to the lowest temperature probed in this experiment, confirming the thermal equilibrium of the electrons in the quantum wires with the apparatus.  The re-entrant negative drag signal disappears at $T \sim 1.2 K$, which is consistent with the system leaving the mesoscopic regime as the temperature length $L_{T}=\hbar v_{F}/k_{B}T$ is lowered from $\sim 5.5 \mu m$ at 0.33 K to $\sim 1.5 \mu m$ at 1.2 K, and becomes shorter than the system size.  Fig. 4\textbf{c} and 4\textbf{d} show the linearity of the drag voltage with drive current (for small enough drive voltages, \emph{i.e.} empirically for $eV_{drive}/K_{b} \lesssim 3 K$) and the probe symmetry of the drag signal, confirming that the signals observed are consistent with Coulomb drag. For wires with a similar subband occupancy presented in Fig. 4\textbf{a}, the drag effect is $ \sim 2\%$ of the drive voltage value.  However, in wires with significantly different subband occupancy, this effect can be as large as $25\%$.

Coulomb drag between nanoelectronic circuits will become increasingly important as nano-circuitry becomes coupled by proximity.  As nanostructure cross-sections become comparable to the 3D screening length, the effective 1D screening length is expected to become large\cite{Screening}.  Using typical doping values for silicon nanowires \cite{Sinanowire}, the bulk Thomas-Fermi screening length $\lambda_{f} =  \sqrt{\frac{\epsilon E_{f}}{6 \pi e^{2} n}}$, where $e$ is the electron change, $n$ is the electron density, $\epsilon$ is the silicon dielectric constant and $E_{f}$ is the Fermi energy, is estimated to be $\sim$ 4 nm.  Therefore, as nanowire diameter approaches this length scale, the previously screened Coulomb interactions will induce Coulomb drag signals in circuit elements located in close proximity.  This drag effect is found to be as large as $25 \%$ of the drive voltage value, or up to $10 \mu$V, for the structures presented in this Letter, which is far from negligible. An understanding of one-dimensional Coulomb drag phenomenon in model systems such as quantum wires will ultimately prove to be an essential asset to understand the coupling between independently addressed conductors at the nanoscale, for example coupled nanowires for nanoprocessing \cite{nanoprocessors}.
\\
\\


\large{\bf Methods}
\\
\\
\normalsize{{\bf Device fabrication}
\\
 The wires are patterned on a n-doped GaAs/AlGaAs electron bilayer heterostructure where two 18 nm wide quantum wells are separated by a 15 nm wide Al$_{0.3}$Ga$_{0.7}$As barrier.  After a mesa-structure is wet-etched using phosphoric acid into the double quantum well heterostrocture, Ge-Au-Ni-Au ohmic contacts are deposited on the structure (Fig. 1\textbf{a}).  Following an annealing at $ 420^{\circ}{\rm C}$ for 60 seconds, a set of two Ti-Au split-gates, consisting of a T-shaped pinch-off gate and of a plunger gate, is defined on the surface of the heterostructure.  The off-mesa patterning is defined using photo-lithography while electron-beam lithography is used to pattern the gates on-mesa (Fig 1\textbf{b}).  The thickness of the gates is 160 nm off-mesa and 60 nm on-mesa.  A set of four alignment marks is also patterned simultaneously to the patterning of the e-beam lithography defined top gates.  These marks are used to align the lower gates to the upper ones. Once the upper side processing is completed, bare GaAs is glued on top of the substrate and the sample is flipped, mechanically lapped and chemically etched until the lower 2DEG is only $\sim$ 150 nm away from the lower surface (which is now on top of the device, as show in Fig. 1\textbf{c}), following an EBASE technique \cite{EBASE}.  Two stop-etch layers are incorporated in the original heterostructure : a larger AlGaAs stop-etch layer and a thinner GaAs stop-etch layer.  The AlGaAs stop-etch layer purpose is to flatten out the unevenness arising from the lapping process during the subsequent citric wet-etching.  Indeed, the citric acid etch rate is greatly reduced in AlGaAs compared to GaAs, allowing to smooth the surface of the device after the mechanical lapping.  After the citric etch, the remainder of the AlGaAs stop-etch layer is etched using hydrofluoric acid, leaving only the thin GaAs stop-etc layer, which is grown to prevent over-etching during the hydrofluoric etch.  To ensure that no off-mesa leakage occurs between the upper and the lower gates, a thin 60 nm layer of Al$_{2}$O$_{3}$ is deposited on the top of the device using atomic layer deposition. Using phosphoric acid, vias are then etched through the surface to enable electrical connection to the ohmic contacts and to the upper split gates on the buried surface of the device.  Finally, using a combination of photolithography and electron-beam lithography, another set of two Ti-Au spit gates is defined on the lower-side of the sample and aligned with the upper gates using the previously deposited alignment marks buried underneath the surface.  It is possible to observe these marks using a SEM or an e-beam lithography tool with an accelerated voltage greater than or equal to 30 keV, and therefore precisely align the lower and the upper gates.  The end result is presented in Fig. 1\textbf{d}.
 \\
 \\
{\bf Device operation}
 \\
 The pinch-off gates are first adjusted such that they principally deplete the 2DEG closest to them.  While each pinch-off gate can deplete both 2DEGs for sufficiently large applied negative voltage, a 0.3 V (0.15 V) wide plateau (where the conductance across the device is roughly constant) is observed when sweeping the upper (lower) pinch-off gate.  On this plateau, the 2DEG closest to the gate is fully depleted whereas the other one is only partially depleted.  For the device presented in this Letter, the lower gates create a larger partial depletion than the upper gates, causing the contact resistance to the upper wire to be larger than the contact resistance of the lower wire.  The positioning within the plateaus is adjusted such that the tunneling resistance between both layers is larger than 25 $M\Omega$.  In such experimental configuration, there is minimal tunneling between the upper and the lower layer contacts.  Indeed, the depletion mechanism of the pinch-off gates results in a coupling of each side of the device to a single layer, allowing simultaneous and independent measurement of both layers, unlike in the device presented by Bielejec \emph{et al.}\cite{Ed1}  Subsequently, adjusting both the lower and the upper plunger gate voltages allows for the independent tuning of the subband occupancy in each independent wire.
 \\
 \\
{\bf Device characterization}
\\
 Measurements performed on the sample post processing with the split-gates grounded yielded an electronic density of 1.1 (1.4) $\times 10^{11}$ cm$^{-2}$ for the upper (lower) 2DEG, and a combined mobility of $4.0 \times 10^{5}$ cm$^{2}$ / V$\cdot$ s. Transport measurements on individual quantum wires were performed in a $^{3}He$ refrigerator at a temperature of 330 mK using a constant 50$\mu$V excitation at 9 Hz in the lower wire and at 13 Hz in the upper wire in a two-contact configuration.  The Coulomb drag measurements were performed in a constant current mode where 4.5 nA at 9 Hz was sent through the drive wire.  In this configuration, the out-of-phase current was always much smaller than the in-phase current.}
\\
\\

\begin{small}

\end{small}

\vspace*{10mm}

\large{{\bf Acknowledgment}}
\\
\normalsize{We acknowledge the outstanding technical assistance of Denise Tibbetts and James Hedberg. We thank Aashish Clerk for inspiring discussions.  This work has been supported by the Division of Materials Sciences and Engineering, Office of Basic Energy Sciences, US Department of Energy.  This work was performed, in part, at the Center for Integrated Nanotechnologies, a U.S. DOE, Office of Basic Energy Sciences user facility.  Sandia National Laboratories is a multi-program laboratory managed and operated by Sandia Corporation, a wholly owned subsidiary of Lockheed Martin Corporation, for the U.S. Department of Energy's National Nuclear Security Administration under contract DE-AC04-94AL85000.  The authors also acknowledge the financial support from the Natural Sciences and Engineering Research Council of Canada (NSERC), CIFAR, and from the FQNRT (Qu\'ebec)}
\\
\\
\end{document}